\documentstyle[prd,aps,epsf,psfig,floats,preprint]{revtex}

\def\beq{\begin{equation}}
\def\eeq{\end{equation}}
\def\ber{\begin{eqnarray}}
\def\eer{\end{eqnarray}}
\def\l{\Lambda}
\def \lleq {\lower0.9ex\hbox{ $\buildrel < \over \sim$} ~}
\def \ggeq {\lower0.9ex\hbox{ $\buildrel > \over \sim$} ~}

\def\prd{{Phys.\@ Rev.\@ D\ }}

\def\etal{{\it et al.}}
\def\ie {{\it ie}}
\def\n {\noindent}

\begin{document}

\draft
\title{Supernova Constraints on Braneworld Dark Energy}

\author{Ujjaini Alam$^a$ and Varun Sahni$^a$ }
\address{$^a$ Inter-University Centre for Astronomy and Astrophysics,
Post Bag 4, Ganeshkhind, Pune 411~007, India\\
}
\maketitle

\begin{abstract}
  Braneworld models of dark energy are examined in the light of
  observations of high redshift type Ia supernovae.  Braneworld models
  admit several novel and even exotic possibilities which include: (i)
  The effective equation of state of dark energy can be more negative
  than in LCDM: $w \leq -1$; (ii) A class of braneworld models can
  encounter a `quiescent' future singularity at which the energy
  density and the Hubble parameter remain well behaved, but higher
  derivatives of the expansion factor ($\stackrel{..}{a}$,
  $\stackrel{...}{a}$ etc.) diverge when the future singularity is
  reached; (iii) The current acceleration of the universe is a {\em
    transient feature} in a class of models in which dark energy
  `disappears' after a certain time, giving rise to a matter dominated
  universe in the future.  Since horizons are absent in such a
  space-time, a braneworld model with {\em transient acceleration}
  might help reconcile current supernova-based observations of an
  accelerating universe with the demands of string/M-theory.  A
  maximum likelihood analysis reveals that braneworld models satisfy
  the stringent demands imposed by high redshift supernovae and a
  large region in parameter space agrees marginally better with
  current observations than LCDM. For instance, models with $w < -1 \ 
  ( \ > -1 )$ provide better agreement with data than LCDM for
  $\Omega_m \ggeq 0.3 \ ( \ \lleq 0.25 )$.
\end{abstract}


\bigskip 

\section{Introduction}

The idea that the universe may have more than three spatial dimensions
first appeared in the works of Kaluza and Klein over 70 years ago.
The notion of extra dimensions gradually became popular because of the
hope that one might succeed in relating the gauge symmetries of
particle physics to the isometries of a compact higher dimensional
manifold \cite{kk}.  All such models, however, assumed that the
compactification scale is small ($l \sim l_{\rm Pl} \sim 10^{-33}$ cm)
hence unobservable. More recently, the many variants of superstring
theory allow for the possibility that at least some of the extra
dimensions of nature may be macroscopic. For example, the eleven
-dimensional supergravity model of Horava and Witten \cite{hw} assumes
that ordinary matter fields are confined to a submanifold (brane)
which is embedded in a higher dimensional space (bulk).  In an
important recent development, Randall and Sundrum (RS) examined a
simplified variant of this model consisting of a three dimensional
brane embedded in a four dimensional anti-de Sitter (AdS) bulk
\cite{rs}.  Their results showed that gravitational excitations are
confined close to the brane giving rise to the familiar $1/r^2$ law of
gravity, and suggesting that one could identify the brane as our
observable universe.

Subsequently the RS ansatz was generalised to incorporate both
expanding FRW-type models \cite{brane} and anisotropic space-times
\cite{anis}.  Of particular importance to the present study is the
observation that simple extensions of the RS scenario can give rise to
a universe which is {\em accelerating}, in agreement with studies of
high redshift supernovae \cite{dgp,ss02a}. It is now well established
that high redshift type Ia supernovae appear fainter than expected in
a spatially flat matter dominated (Einstein-de Sitter) universe
\cite{riess,perl}.  One way of explaining this discrepancy is to
postulate that the universe is filled with a smooth component carrying
large negative pressure (dark energy). Although several possible
candidates for dark energy have been suggested (the cosmological
constant, quintessence etc.)  none is entirely problem free (see
\cite{ss00,sahni02} for recent reviews).

In this paper we shall focus on a new form of dark energy based on the
braneworld model examined in \cite{ss02a,ss02b} (see also
\cite{CH,Shtanov1}).  Braneworld models of dark energy have
interesting new properties including the fact that, depending upon the
form of bulk-brane embedding, the effective equation of state of dark
energy can be $w \geq -1$ or $w \leq -1$.  In addition, for an
appropriate parameter choice, the acceleration of the universe can be
a transient phenomenon, thus helping reconcile high-$z$ supernova
observations of an accelerating universe with the requirements of
string/M-theory.

\section{Dark Energy from Braneworld models}

The equations of motion governing the braneworld can be derived from
the action \cite{CH,Shtanov1} 
\beq \label{action} 
S = M^3 \left[\int_{\rm bulk} \left( R_5 - 2 \l_{\rm b} \right) - 2
  \int_{\rm brane} K \right]
+ \int_{\rm brane} \left( m^2 R_4 - 2 \sigma \right) + \int_{\rm
  brane} L \left( h_{\alpha\beta}, \phi \right) \, .  
\eeq 
Here, $R_5$ is the scalar curvature of the metric $g_{ab}$ in the
five-dimensional bulk, and $R_4$ is the scalar curvature of the
induced metric $h_{\alpha\beta}$ on the brane. The quantity $K =
K_{\alpha\beta} h^{\alpha\beta}$ is the trace of the extrinsic
curvature $K_{\alpha\beta}$ on the brane defined with respect to its
{\em inner\/} normal. $L (h_{\alpha\beta}, \phi)$ is the
four-dimensional matter field Lagrangian, $M$ and $m$ denote,
respectively, the five-dimensional and four-dimensional Planck masses,
$\l_{\rm b}$ is the bulk cosmological constant, and $\sigma$ is the
brane tension.  Integrations in (\ref{action}) are performed with
respect to the natural volume elements on the bulk and brane.

The presence of the brane curvature term $m^2\int_{\rm brane}R_4$ in
(\ref{action}) introduces an important length scale into the problem $
l = 2m^2/M^3$. On short length scales $r \ll l$ (early times) one
recovers general relativity, whereas on large length scales $r \gg l$
(late times) brane-specific effects begin to play an important role,
leading to the acceleration of the universe at late times
\cite{dgp,DDG,ss02a}).

The cosmological evolution of the braneworld is described by the
Hubble parameter

\beq \label{eq:hubble} 
H^2 + {\kappa \over a^2} = {\rho + \sigma \over
  3 m^2} + \underline{{2 \over l^2} \left[1 \pm \sqrt{1 + l^2
      \left({\rho + \sigma \over 3 m^2} - {\l_{\rm b} \over 6} - {C
          \over a^4} \right)} \right]} \, .  
\eeq
The two signs in (\ref{eq:hubble}) correspond to the two separate ways
in which the brane can be embedded in the higher dimensional bulk.
The underlined term in (\ref{eq:hubble}) makes braneworld models
different from standard FRW cosmology.  The limiting case of our model
$m=0$ corresponds to the well known FRW generalisation of the RS
scenario \cite{brane} 
\beq\label{eq:cosmolim} 
H^2 + {\kappa \over a^2} 
= {\l_{\rm b} \over 6} + {C \over a^4} + {(\rho + \sigma)^2 \over 9
  M^6}\, .  
\eeq 
In this case, braneworld evolution departs from the standard FRW law
at early times when $\rho/\sigma \gg 1$. However, as remarked earlier,
braneworld models described by (\ref{eq:hubble}) depart from FRW
behaviour at {\em late times}, a property that opens radically new
possibilities for the present and future state of our universe.

Braneworld models fall into three main categories \cite{ss02a}:

\begin{itemize}
\item{1. {\bf BRANE1} (B1):} The Hubble parameter is given by
\beq \label{eq:hubble1}
{H^2(z) \over H_0^2} = \Omega_{\rm m} (1\!+\!z)^3 + \Omega_\kappa (1\!+\!z)^2 +
\Omega_\sigma 
+\underline{2 \Omega_l - 2 \sqrt{\Omega_l}\,
\sqrt{\Omega_{\rm m} (1\!+\!z )^3 + \Omega_\sigma + \Omega_l +
\Omega_{\l_{\rm b}}}} \, , 
\eeq
where $z = a_0/a(t)-1$ is the cosmological redshift, while
\beq \label{eq:omegas}
\Omega_{\rm m} =  {\rho_0 \over 3 m^2 H_0^2} \, , \quad \Omega_\kappa = -
{\kappa \over a_0^2 H_0^2} \, , \quad \Omega_\sigma = {\sigma \over 3 m^2
H_0^2} \, , 
\quad \Omega_l = {1 \over l^2 H_0^2} \, , \quad
\Omega_{\l_{\rm b}} = - {\l_{\rm b} \over 6 H_0^2} \, ,
\eeq
are dimensionless parameters whose values must be determined from
observations (the subscript `{\small 0}' refers to their current
value).  $\Omega_\sigma$ is determined by the constraint relation
\beq \label{eq:omega1}
\Omega_{\rm m} + \Omega_\kappa + \Omega_\sigma - \underline{2
\sqrt{\Omega_l}\, \sqrt{1 - \Omega_\kappa + \Omega_{\l_{\rm b}}}} = 1.
\eeq

\item{2. {\bf BRANE2} (B2):} The Hubble parameter is given by
\beq \label{eq:hubble2}
{H^2(z) \over H_0^2} = \Omega_{\rm m} (1\!+\!z)^3 + \Omega_\kappa (1\!+\!z)^2 +
\Omega_\sigma \nonumber\\ 
+ \underline{2 \Omega_l + 2 \sqrt{\Omega_l}\,
\sqrt{\Omega_{\rm m} (1\!+\!z )^3 + \Omega_\sigma + \Omega_l +
\Omega_{\l_{\rm b}}}} \, , \nonumber\\
\eeq
where $\Omega_l < 1 + \Omega_{\l_{\rm b}}$
and $\Omega_\sigma$ is determined from
\beq \label{eq:omega2}
\Omega_{\rm m} + \Omega_\kappa + \Omega_\sigma + \underline{2
\sqrt{\Omega_l}\, \sqrt{1 - \Omega_\kappa + \Omega_{\l_{\rm b}}}} = 1.
\eeq

The two models BRANE1 and BRANE2 are complementary and reflect the two
distinct ways in which the brane can be embedded in the bulk
\cite{ss02a}.  Clearly, by setting $\Omega_l = 0$ the underlined terms
in (\ref{eq:hubble1}), (\ref{eq:omega1}), (\ref{eq:hubble2}),
(\ref{eq:omega2}) vanish and we recover the LCDM model in both cases.
The underlined terms in (\ref{eq:hubble1}) \& (\ref{eq:hubble2}) are
caused by braneworld effects and give rise to an important general
result, namely 
\beq
H(z)\bigg\vert_{\rm B1} \leq H(z)\bigg\vert_{\rm LCDM}, ~~~~ 
H(z)\bigg\vert_{\rm LCDM} \leq H(z)\bigg\vert_{\rm B2} \leq
H(z)\bigg\vert_{\rm SCDM},
\label{eq:hubble_ineq}
\eeq
\ie ~~the universe expands at a faster (slower) rate than LCDM in
BRANE2 (BRANE1). Since most cosmological observables involve $H(z)$
either directly or indirectly, braneworld models can exhibit
properties which can be quite distinct from those of either LCDM or
SCDM.

\item{3. {\bf Disappearing dark energy} (DDE):} The Hubble parameter
  is given by (\ref{eq:hubble2}), and the cosmological parameters
  $\Omega_{\rm m}, \Omega_\sigma, \Omega_l, \Omega_{\l_{\rm b}}$ are
  constrained to satisfy the following relations \cite{ss02a}:

\beq\label{eq:constr}
\Omega_l \leq \Omega_{\l_{\rm b}} \Rightarrow
\Omega_{\l_{\rm b}} \ge \frac{\left(1 - \Omega_{\rm m} \right)^2}{4
\Omega_{\rm m}} \,,
\eeq

\beq \label{eq:omegl}
\Omega_{\rm m} + 2 \sqrt{\Omega_l} \left( \sqrt{1 + \Omega_{\l_{\rm
b}}} - \sqrt{\Omega_{\l_{\rm b}}} \right) = 1,
\eeq

\beq \label{eq:omegsig}
\Omega_\sigma = - 2\sqrt{\Omega_l\Omega_{\l_{\rm b}}}.
\eeq
(We make the assumption that the universe is spatially flat, so that
$\Omega_\kappa = 0$.)

\end{itemize}

From (\ref{eq:hubble1}) \& (\ref{eq:hubble2}) it is easy to see that
all braneworld models approach the standard matter dominated universe
at early times [with a small correction term $\sim (1+z)^{3/2}$]. At
late times the behaviour of the braneworld can differ from both LCDM
and SCDM.  This feature makes braneworld models testable and allows
the braneworld scenario to provide a new explanation for the
observational discovery of dark energy.

The braneworld models described above provide a common platform for
understanding the properties of dark energy. For instance the
expression for the Hubble parameter in these models (\ref{eq:hubble1})
\& (\ref{eq:hubble2}) allows us to explicitly determine key
cosmological quantities including:

\begin{itemize}

\item The luminosity distance $d_L(z)$:

\beq
{d_L(z) \over 1 + z} = c\int_0^z {dz' \over H(z')} \, ,
\label{eq:lumdis2}
\eeq

\item the angular-size distance

\beq
d_A(z) = \frac{c}{1+z}\int_0^z {dz' \over H(z')} \, ,
\label{eq:angdis}
\eeq

\item the deceleration parameter:

\beq \label{eq:decel}
q(z) = \frac{H'(z)}{H(z)} (1+z) - 1 \, ,
\eeq

\item the effective equation of state of dark energy:
\beq \label{eq:state0}
w(z) = {2 q(z) - 1 \over 3 \left[ 1 - \Omega_{\rm m}(z) \right] } \, ,
\eeq

\item the age of the universe:

\beq\label{eq:age}
t(z) = \int_z^\infty \frac{dz'}{(1+z') H(z')} \, ,
\eeq

\item the `statefinder pair' \cite{sssa}:

\ber
r &=& \frac{\stackrel{...}{a}}{a H^3} \equiv  1 + \left\lbrack \frac{H''}{H} + \left (\frac{H'}{H}\right )^2
\right\rbrack
(1+z)^2 - 2\frac{H'}{H}(1+z),\nonumber\\
s &=& \frac{r - 1}{3(q - 1/2)},
\label{eq:state}
\eer

\item the product $d_A(z)H(z)$, which is used in the
Alcock-Paczynski anisotropy test \cite{alcock},

\item the product $d_A^2(z)H^{-1}(z)$, which plays a key role in the
  volume-redshift test \cite{davis}.
\end{itemize}

While providing a common basis for the existence of dark energy,
braneworld models B1, B2, DDE, have important properties and
attributes which serve to distinguish these models both from each
other, and from alternate models of dark energy such as LCDM \&
quintessence. For instance, the luminosity distance in BRANE1 {\em can
  be larger} than the luminosity distance in LCDM: $d_L|_{B1} \geq
d_L|_{LCDM}$. This follows from (\ref{eq:hubble_ineq}) and leads to an
important result. Namely, using (\ref{eq:state0}) we find the current
value of the effective equation of state of dark energy:

\beq
w_0 = {2 q_0 - 1 \over 3 \left( 1 - \Omega_{\rm
      m} \right)} = - 1 - {\Omega_{\rm m} \over 1 - \Omega_{\rm m}} \,
{\sqrt{\Omega_l \over \Omega_{\rm m} + \Omega_\sigma + \Omega_l +
    \Omega_{\l_{\rm b}}}} \, ,  
\eeq 
from where we see that $w_0 \leq -1$. BRANE1 therefore has the
remarkable property that the effective equation of state of dark
energy can be more negative than that associated with a cosmological
constant ($w = -1$).  This feature distinguishes B1 models from LCDM
as well as from scalar field based quintessence models.

BRANE2 has the opposite property, namely $d_L|_{SCDM} < d_L|_{B2} \leq
d_L|_{LCDM}$, which translates into

\ber
-1 \leq w_0 & < & 0, \nonumber\\
{\rm where} \, \,\, w_0 &=& - 1 + {\Omega_{\rm m} \over 1 -
  \Omega_{\rm m}} \, {\sqrt{\Omega_l \over \Omega_{\rm m} +
    \Omega_\sigma + \Omega_l + \Omega_{\l_{\rm b}}}} \, .  
\eer

\begin{figure*}
\centerline{ \psfig{figure=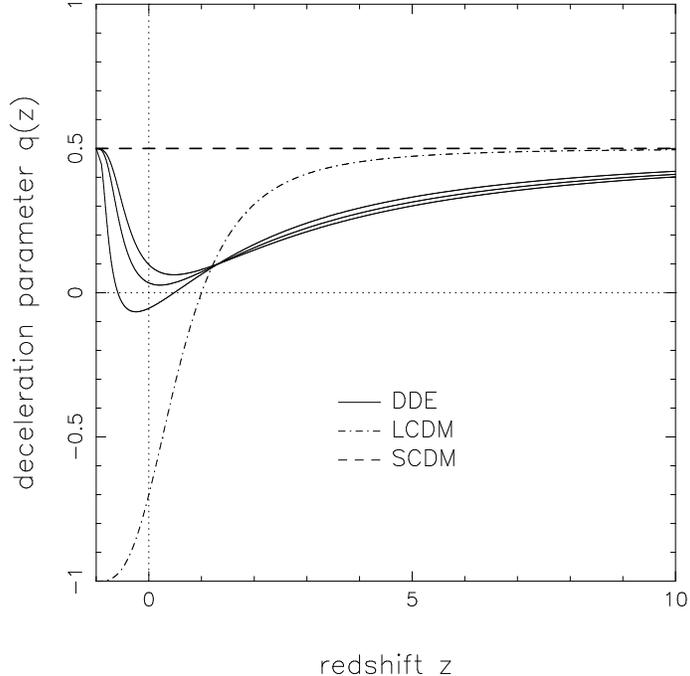,width=0.55\textwidth,angle=0} }
\bigskip
\caption{\small 
  The behaviour of the deceleration parameter with redshift for DDE. 
  Solid curves from top to bottom are DDE models with
  $\Omega_m=0.2$, and $\Omega_{\l_b}=1.0, 1.5, 2.0$ respectively.
  The dashed line is SCDM with $\Omega_m=1$, the dot-dashed curve
  is LCDM with $\Omega_m=0.2$. The vertical dotted line marks the
  present epoch at $z=0$, and the horizontal dotted line marks a $q=0$
  Milne universe. In the DDE models, the universe ceases to accelerate
  and becomes matter-dominated in the future, unlike the LCDM model
  which remains dark energy dominated at all future times.}
\label{fig:dde}
\end{figure*}

We therefore see that B1 \& B2 are complementary models and that the
effective equation of state in B1 (B2) is {\em softer} (stiffer) than
the $w=-1$ typical of LCDM.  Furthermore in both B2 \& DDE, the
current acceleration of the universe -- like earlier matter and
radiation dominated epochs -- can be a {\em transient feature}.  In
the case of DDE, the current accelerating phase will be replaced by a
matter dominated epoch during which $w \simeq 0$ \cite{ss02a}. In Fig.
\ref{fig:dde} we show the behaviour of the deceleration parameter for
this class of models. It is seen that braneworld dark energy
disappears in the future and the universe becomes matter dominated as
$z \to -1$. It is easy to show that the present value of the effective
equation of state of dark energy in DDE is given by

\beq\label{eq:dde_w} 
w_0 = -\frac{1}{2}\frac{\sqrt{1+\Omega_{\l_b}} -
  \sqrt{\Omega_{\l_b}}} {\sqrt{1+\Omega_{\l_b}} - \sqrt{\Omega_l}}.
\eeq 

\begin{figure*}
\centerline{ \psfig{figure=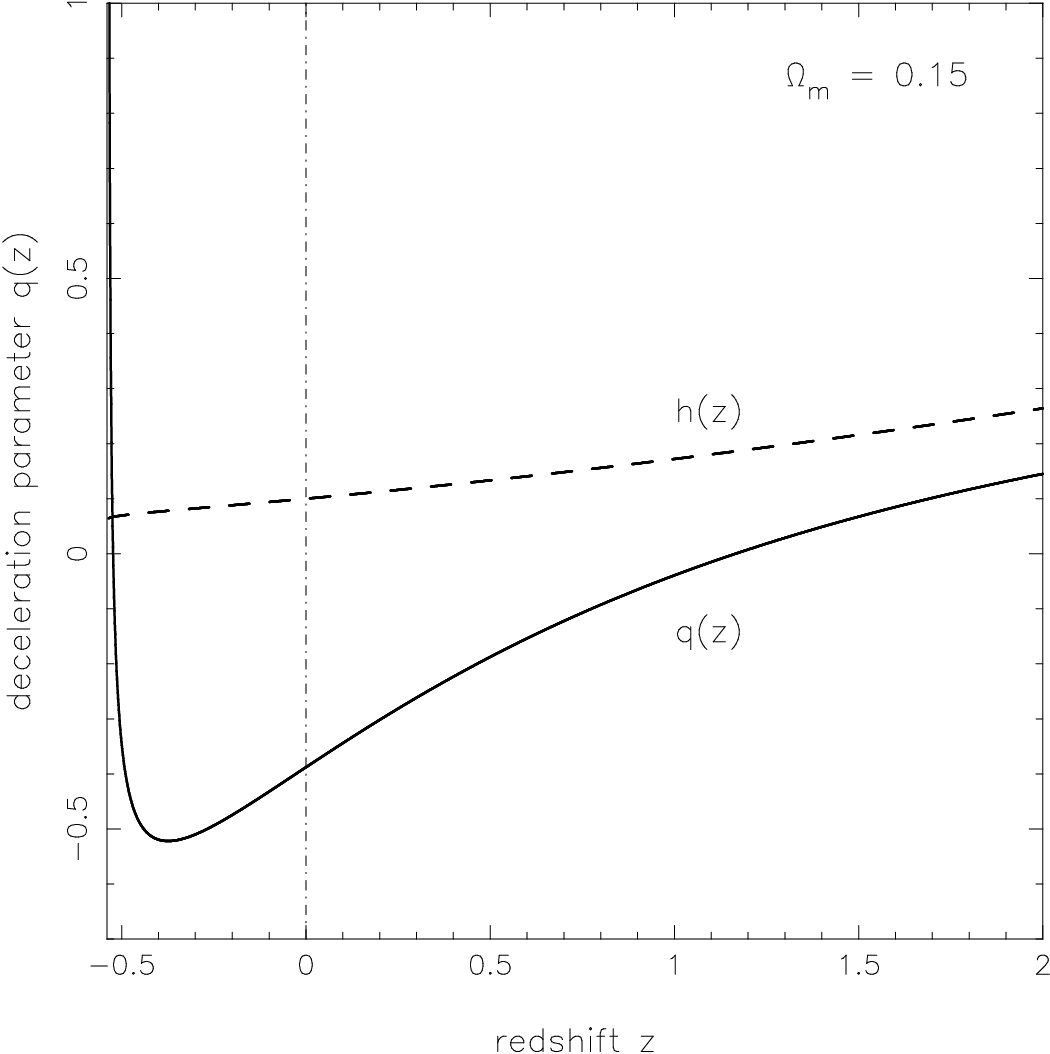,width=0.5\textwidth,angle=0} }
\bigskip
\caption{\small The BRANE2 universe can encounter a singularity
  lying in the {\em future} as demonstrated in this figure.  The
  deceleration parameter becomes infinite as the singularity is
  approached while the Hubble parameter (and the density, pressure)
  remain finite.  The vertical dot-dashed line corresponds to the
  present epoch $z=0$ while the dashed line represents the
  dimensionless Hubble parameter. The solid line shows the
  deceleration parameter $q(z)$.  The model parameters are $\Omega_m =
  0.15, ~\Omega_l = 0.4$.  Although permitted by supernovae
  observations this particular model appears to be disfavored by
  clustering bounds on $\Omega_m$ (see Fig. \ref{fig:contour2}).}
\label{fig:sing}
\end{figure*}

\n Since $\Omega_l \leq \Omega_{\l_b}$ we find that $w_0 \geq -0.5$.
Model B2 permits more exotic possibilities.  In a subclass of models
acceleration gives way to an epoch during which the universe
decelerates at an increasingly rapid rate \cite{ss02b}.  The expansion
of the universe culminates in a `quiescent singularity' which is
distinguished from conventional general relativistic singularities by
the fact that the energy density and Hubble parameter {\em remain
  finite}, while higher derivatives of the scale factor
($\stackrel{..}{a}$, $\stackrel{...}{a}$ etc.) diverge, when the
`future singularity' is reached (Fig. \ref{fig:sing}).  (The future
singularity, measured from the present epoch, is reached in a few
billion years for most B2 type braneworld models \cite{ss02b}.)

Since neither B2 nor DDE possess an event horizon both can
successfully reconcile the demands of string/M-theory with a universe
which is currently accelerating \cite{horizon}.

\section{Methodology and Results}

We constrain the parameter space of braneworld cosmology by ensuring
that our cosmological models provide a good fit to Type Ia supernova
data. For this purpose we use the 54 SNe Ia from the primary `fit C'
of the Supernova Cosmology Project, which includes 16 low redshift
Calan-Tololo SNe \cite{perl}. Fit C is a subsample of a total of 60
SNe of which six are excluded as outliers: two low redshift SNe due to
suspected reddening and four high redshift SNe of which two are
excluded due to atypical light curves and two because of reddening.
The measured quantity in this data, the bolometric magnitude $m_B$, is
related to the luminosity distance and therefore the cosmological
parameters by the following equation

\beq\label{eq:mz}
 m_B = {\cal M}+ 5 \ {\rm log}_{10} D_L(z;\Omega_m, \Omega_l,
 \Omega_{\l_b})\,\,,
\eeq

\n where $D_L=H_0 d_L$ is the Hubble-parameter-free luminosity
distance and $ {\cal M}= M_B + 25 -5 \ {\rm log}_{10} H_0$ is the
Hubble-parameter-free absolute magnitude.  We shall assume that the
SNe measurements come with uncorrelated Gaussian errors in which case
the likelihood function is given by the chi-squared distribution with
$N-n$ degrees of freedom ${\cal L} \propto \exp{(-\chi^2/2)}$. ( In
our case $N = 54$, and $n=4$: B1 and B2 estimate four parameters each,
and DDE estimates three parameters with one constraint, Eq
(\ref{eq:constr}) lowering the degrees of freedom)

 The $\chi^2$-statistic is defined as
\beq
\chi^2=\sum_{i=1}^{54} \left (\frac{m_i^{\rm eff}-m(z_i)}{\sigma_{m_i}}\right)^2 \,\,,
\eeq 
\n where $m_i^{\rm eff}$ is the effective B-band magnitude of the i-th
supernova obtained after correcting the observed magnitude at redshift
$z$ for the supernova width-luminosity relation, $\sigma_{m_i}$ is the
error in magnitude at redshift $z$, and $m(z_i)$ is the apparent
magnitude of the i-th supernova in the braneworld model.

\begin{figure*}
\centerline{\mbox{BRANE1 (B1)}}
\vspace{0.1in}
\centerline{ \psfig{figure=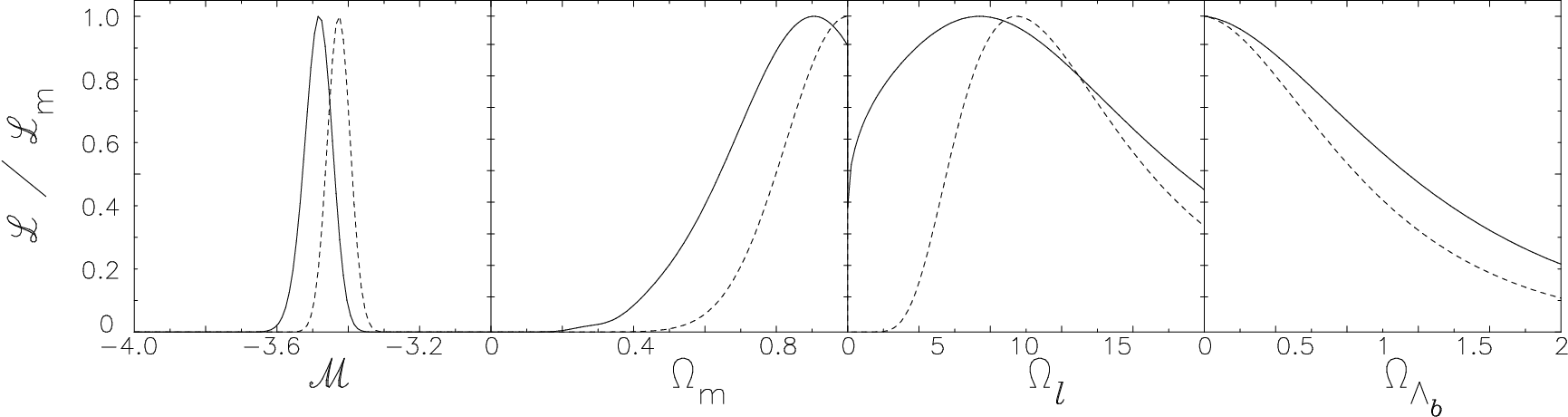,width=1.0\textwidth,angle=0} }
\centerline{\mbox{BRANE2 (B2)}}
\vspace{0.1in}
\centerline{ \psfig{figure=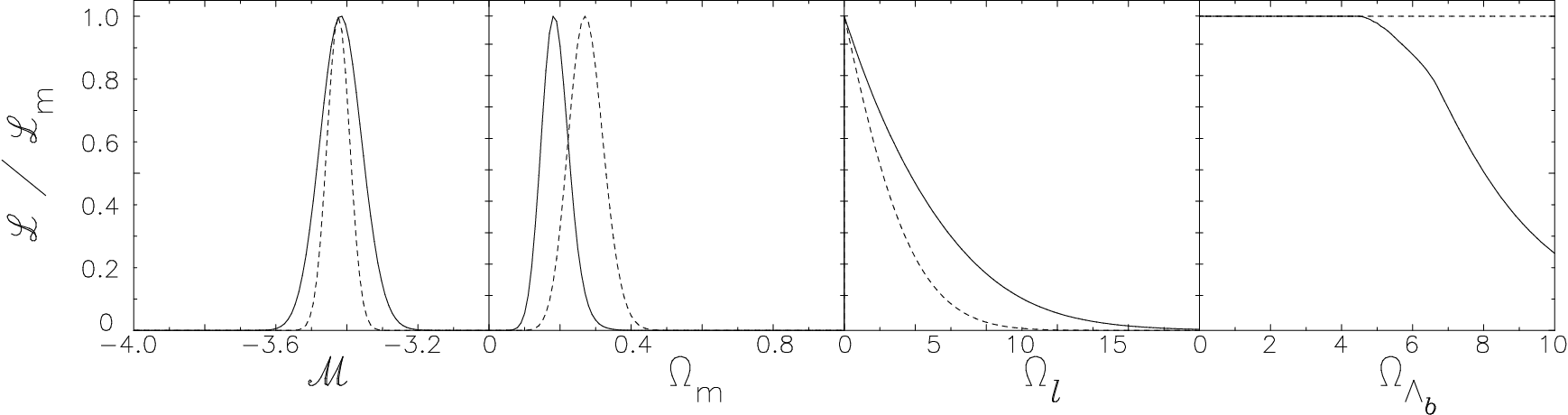,width=1.0\textwidth,angle=0} }
\centerline{\mbox{Disappearing Dark Energy (DDE)}}
\vspace{0.1in}
\centerline{ \psfig{figure=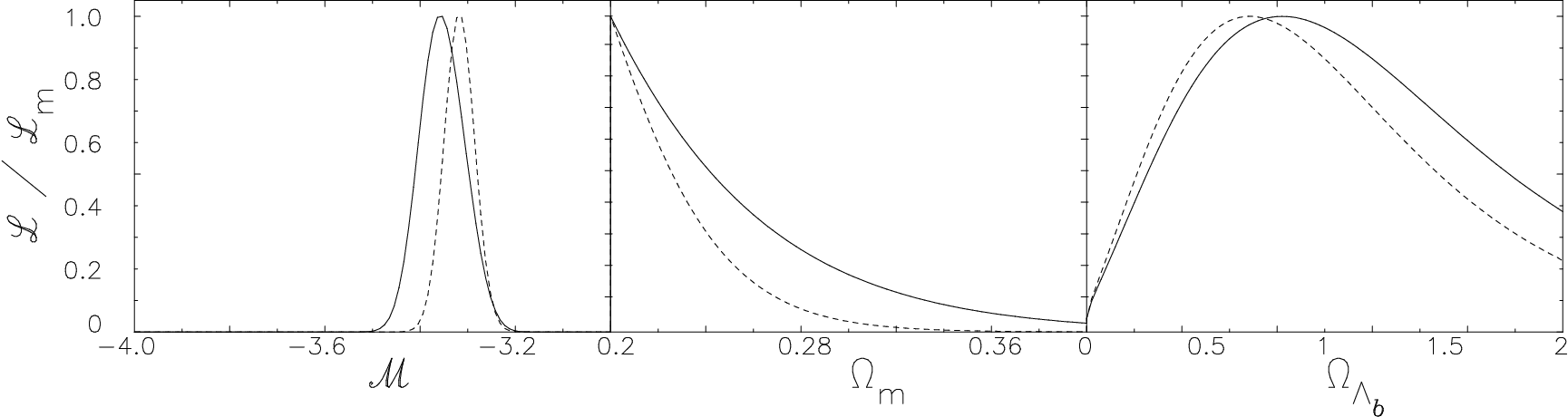,width=1.0\textwidth,angle=0} }
\bigskip
\caption{\small The likelihood function is shown as a function of each of
  the parameters $\Omega_m, \ \Omega_l, \ \Omega_{\l_b}$ and
  ${\cal M}$ after the remaining parameters have been marginalised.
  The top, middle and bottom panels show the likelihood curves for
  B1, B2, and DDE.  The solid lines correspond to the value of the
  likelihood function for a given parameter after it has been
  marginalised over all other parameters, while the dashed lines show
  the likelihood function evaluated by fixing the other parameters at
  their maximum likelihood value. In all three cases the likelihood
  function is normalised with the maximum likelihood value set to
  unity.} \label{fig:likelihood}
\end{figure*}

For BRANE1 and BRANE2, the parameters to be estimated are ${\cal M}$,
$\Omega_m, \ \Omega_l, \ \Omega_{\l_b}$ ($\Omega_{\sigma}$ is
calculated from Eqs (\ref{eq:omega1}) and (\ref{eq:omega2})
respectively for B1 \& B2).  For our purposes the quantity ${\cal M}$
is a statistical nuisance parameter, and we marginalise over it
assuming no prior knowledge to get the three-dimensional probability
distribution in the $(\Omega_m, \Omega_l, \Omega_{\l_b})$ space:
$P(\Omega_m, \Omega_l, \Omega_{\l_b})=\int P(\Omega_m, \Omega_l,
\Omega_{\l_b},{\cal M}) d{\cal M}$.  We perform maximum likelihood
analysis on the system with the priors $0 \leq \Omega_m \leq 1, \ 
\Omega_l \geq 0, \ \Omega_{\l_b} \geq 0$. For B2, we use the added
constraint $\Omega_l \leq 1+\Omega_{\l_b}$. For DDE, the parameters to
be estimated are ${\cal M}$, $\Omega_m, \ \Omega_{\l_b}$.  (The
remaining two parameters $\Omega_l$ and $\Omega_{\sigma}$ are related
to $\Omega_m, ~\Omega_{\l_b}$ through Eqs (\ref{eq:omegl}) and
(\ref{eq:omegsig}) respectively). For this model we use the prior $0
\leq \Omega_m \leq 1$ and the constraint Eq (\ref{eq:constr}) (see
\cite{ss02a} Appendix). For B1 \& B2 the constraint relations
(\ref{eq:omega1}), (\ref{eq:omega2}) combined with
$\Omega_{\kappa}=0$, set the lower bound $\Omega_{\l_b} \geq -1$.
However since $\Omega_{\l_b} \geq 0$ is a more physically appealing
model (it includes anti-de Sitter space (AdS) bulk geometry), we
choose this as a prior for further analysis.  Results for $-1 \leq
\Omega_{\l_b} < 0$ will be presented elsewhere.

\begin{figure*}
\begin{center}
$\begin{array}{c@{\hspace{0.4in}}c}
\multicolumn{1}{l}{\mbox{}} &
        \multicolumn{1}{l}{\mbox{}} \\ [-1.0cm]
\epsfxsize=3.2in
\epsffile{conf1_ml.epsi} &
        \epsfxsize=3.2in
        \epsffile{conf1_ml0.epsi} \\ [0.20cm]
\mbox{\bf (a)} & \mbox{\bf (b)}
\end{array}$
\end{center}
\caption{\small
  Confidence levels at $68.3\%$ (light grey inner contour) $95.4\%$
  (medium grey contour) and $99.73\%$ (dark grey outer contour) are
  shown in the $\Omega_l$-$\Omega_m$ plane for BRANE1. Panel (a)
  represents confidence levels in the $\Omega_l$-$\Omega_m$ plane when
  marginalised over $\Omega_{\l_b}$ and ${\cal M}$, while 
  panel (b) shows the confidence levels marginalised over ${\cal M}$,
  with $\Omega_{\l_b}=0$, the best-fit value. We see that taking
  the best-fit value of $\Omega_{\l_b}=0$ instead of
  marginalising over it does not change the results appreciably. The
  dotted region represents the intersection of the $3\sigma$
  confidence level with the observational constraint $0.2 \leq
  \Omega_m \leq 0.5$. The thick solid line in (b) represents the most
  likely value of $\Omega_l$ if the value of $\Omega_m$ is known
  exactly. We see that the BRANE1 model is in good agreement with SNe
  observations if the value of $\Omega_m$ is moderately high:
  $\Omega_m \ggeq 0.3$.  (It should be noted that the value of the
  five dimensional Planck mass corresponding to $\Omega_l \sim 1$ is
  $M \sim 100$ MeV.)  }
\label{fig:contour1}
\end{figure*}

\begin{figure*}
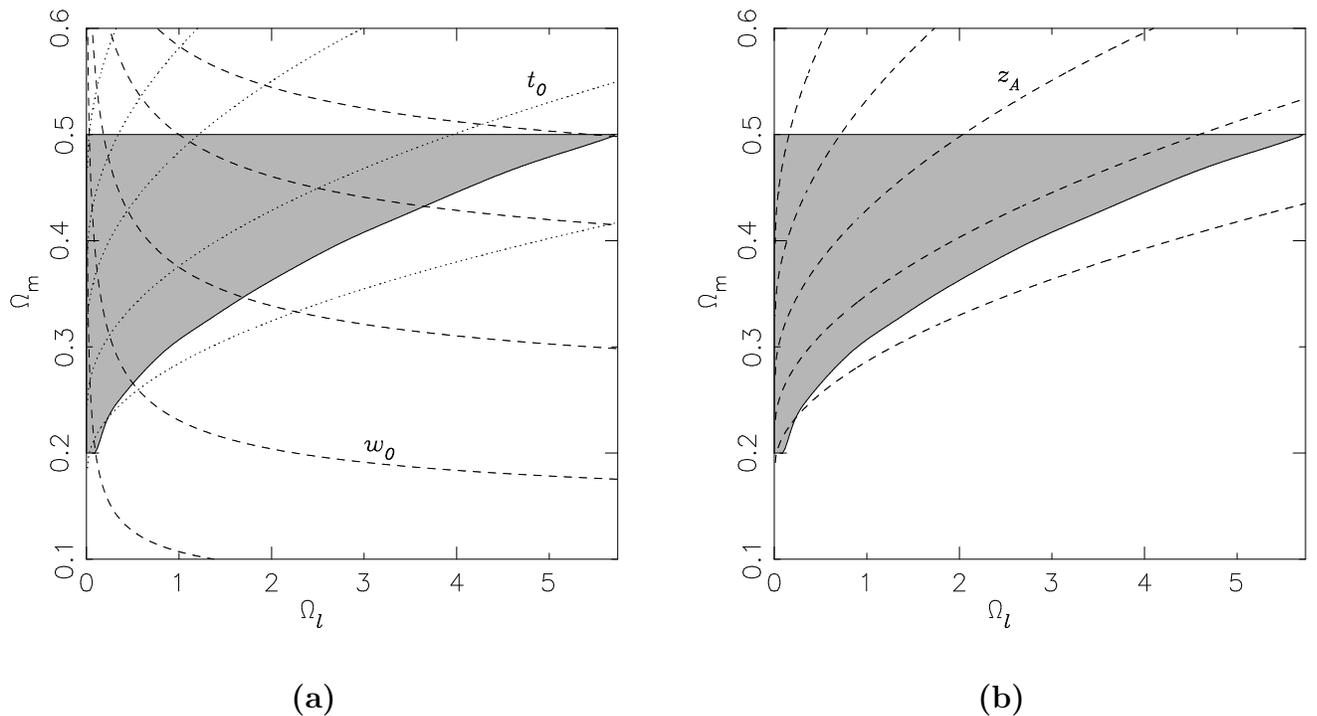

\begin{center}
$\begin{array}{c@{\hspace{0.4in}}c}
\multicolumn{1}{l}{\mbox{}} &
        \multicolumn{1}{l}{\mbox{}} \\ [-1.0cm]
\epsfxsize=3.2in
\epsffile{conf1_0.epsi} &
        \epsfxsize=3.2in
        \epsffile{conf1_z.epsi} \\ [0.20cm]
\mbox{\bf (a)} & \mbox{\bf (b)}
\end{array}$
\end{center}
\caption{\small
  The shaded region represents the intersection of the $3 \sigma$
  confidence region in the $\Omega_m-\Omega_l$ plane for B1 with the
  observational constraint $0.2 \leq \Omega_m \leq 0.5$
  ($\Omega_{\l_b}=0$ is assumed).  In panel (a) we show the values of
  different cosmological quantities determined at the present epoch.
  The dashed lines correspond to the current effective equation of
  state of braneworld dark energy for BRANE1 models: $w_0 = -1.70, \ 
  -1.50, \ -1.30, \ -1.15, \ {\rm and} \ -1.05$ (top to bottom). The
  dotted lines correspond to the current age of the BRANE1 universe:
  $H_0 t_0 = 0.85, \ 0.90, \ 0.95, \ 1.02, \ {\rm and} \ 1.10$ (top to
  bottom). This corresponds to $t_0 ( {\rm Gyrs} ) = 11.8, \ 12.6, \ 
  13.3, \ 14.3 \ {\rm and} \ 15.4$ if $H_0=70 \ {\rm km/sec/Mpc}$.  In
  panel (b) the dashed lines correspond to the epoch, $z_A$, at which
  the braneworld universe (B1) begins to accelerate: $z_A = 0.45, \ 
  0.60, \ 0.75, \ 0.90, \ {\rm and} \ 1.05$ (top to bottom).  }
\label{fig:bounds1}
\end{figure*}

In Fig.  \ref{fig:likelihood} we show the likelihood curve as a
function of $\Omega_m, \ \Omega_l, \ \Omega_{\l_b}$ and ${\cal M}$ for
the braneworld models B1, B2, and DDE.  We see that for all three
cases the likelihood is a sharply peaked function of ${\cal M}$, which
is non-zero over a very limited range. Therefore it seems reasonable
to marginalise over ${\cal M}$ in this range.  Surprisingly, the
likelihood function for BRANE1 peaks at $\Omega_{\l_b}=0$, which is
its best-fit value.  Thus the BRANE1 universe appears to prefer a {\em
  vanishing cosmological constant} in the bulk.  The BRANE2 universe
shows very different behaviour.  In this case the likelihood function
is flat and therefore insensitive to the value of $\Omega_{\l_b}$ over
the wide range $0 \leq \Omega_{\l_b} \lleq 5$. Beyond $\Omega_{\l_b}
\sim 5$ the likelihood function drops steeply.  In Figs
\ref{fig:contour1} and \ref{fig:contour2} we have shown maximum
likelihood contours for B1 and B2 obtained by (a) marginalising over
$\Omega_{\l_b}$, and (b) setting $\Omega_{\l_b}=0$.  The close
similarity between these two figures suggests that the likelihood
contours marginalised over $\Omega_{\l_b}$ do not change the results
appreciably, either for B1 or for B2.

In Fig. \ref{fig:contour1} we show confidence levels in the
$\Omega_m-\Omega_l$ plane for BRANE1.  We find that the BRANE1 model
can be definitely excluded only if the matter density is small
$\Omega_m \lleq 0.2$. For $\Omega_m \ggeq 0.3$ the BRANE1 model agrees
well with supernovae data, with the agreement extending to larger
values of $\Omega_l$ as $\Omega_m$ increases.  Clearly in order to be
able to restrict the braneworld parameters further one needs
additional information about the dark matter density not included in
the supernova observations. Currently there is no firm consensus on
the value of $\Omega_m$. While recent studies of galaxy clustering
indicate $\Omega_m \sim 0.3$ \cite{omegm}, larger values of $\Omega_m$
may be favoured by observations of clusters of galaxies
\cite{borgani,primack}.  In this paper we shall assume the weak
clustering bound $0.2 \lleq \Omega_m \lleq 0.5$ and study the
braneworld models in greater detail in this region. In figure
\ref{fig:contour1} the intersection of the `3$\sigma$' SNe bound and
the bound on $\Omega_m$ is shown as a dotted region.  Interestingly a
large region in BRANE1 parameter space is seen to satisfy both the
supernova constraints as well as the clustering bounds on $\Omega_m$.

Fig. \ref{fig:bounds1} highlights the $\Omega_m-\Omega_l$ region
permitted both by the SNe bound on BRANE1 and by the clustering bound
$0.2 \lleq \Omega_m \lleq 0.5$ for the best-fit value
$\Omega_{\l_b}=0$.  The lines running through the region depict,
respectively, (i) different values of the current effective equation
of state, $w_0$, of dark energy, (ii) the age of the universe, $t_0$,
(iii) the {\em acceleration epoch}, $z_A$, defined as the redshift at
which the universe began to accelerate.  We find that the age of the
universe is constrained to lie in the range $0.85 \lleq t_0H_0 \lleq
1.10$, while the effective equation of state is $-1.70 \lleq w_0 \lleq
-1$ (One should note that $w_0 \leq -1$ in the BRANE1 model) . From
the figure we find that the acceleration of the BRANE1 universe is a
fairly recent phenomenon, which commenced at $0.45 \lleq z_A \lleq
1.05$.

\begin{figure*}
\begin{center}
$\begin{array}{c@{\hspace{0.4in}}c}
\multicolumn{1}{l}{\mbox{}} &
        \multicolumn{1}{l}{\mbox{}} \\ [-1.0cm]
\epsfxsize=3.2in
\epsffile{conf2_ml.epsi} &
        \epsfxsize=3.2in
        \epsffile{conf2_ml0.epsi} \\ [0.20cm]
\mbox{\bf (a)} & \mbox{\bf (b)}
\end{array}$
\end{center}
\caption{\small
  Confidence levels at $68.3\%$ (light grey inner contour) $95.4\%$
  (medium grey contour) and $99.73\%$ (dark grey outer contour) are
  shown in the $\Omega_l$-$\Omega_m$ plane for the BRANE2 model. The
  figure (a) represents confidence levels in the $\Omega_l$-$\Omega_m$
  plane when marginalised over $\Omega_{\l_b}$ and ${\cal M}$,
  while the figure (b) shows the confidence levels marginalised over
  ${\cal M}$, with $\Omega_{\l_b}=0$. We see that fixing the
  value of $\Omega_{\l_b}$ to zero instead of marginalising over
  it has negligible effect on the results. The dotted region
  represents the intersection of the $3\sigma$ confidence level with
  the observational constraint $0.2 \leq \Omega_m \leq 0.5$. The thick
  solid line in (b) represents the most likely value of $\Omega_l$ if
  the value of $\Omega_m$ is known exactly. The region to the right of
  the dotted line in (b) represents B2 universes which will encounter
  future `quiescent' singularities. We see that the BRANE2 model is
  in good agreement with SNe observations for lower values of
  $\Omega_m \lleq 0.30$.  }
\label{fig:contour2}
\end{figure*}

\begin{figure*}
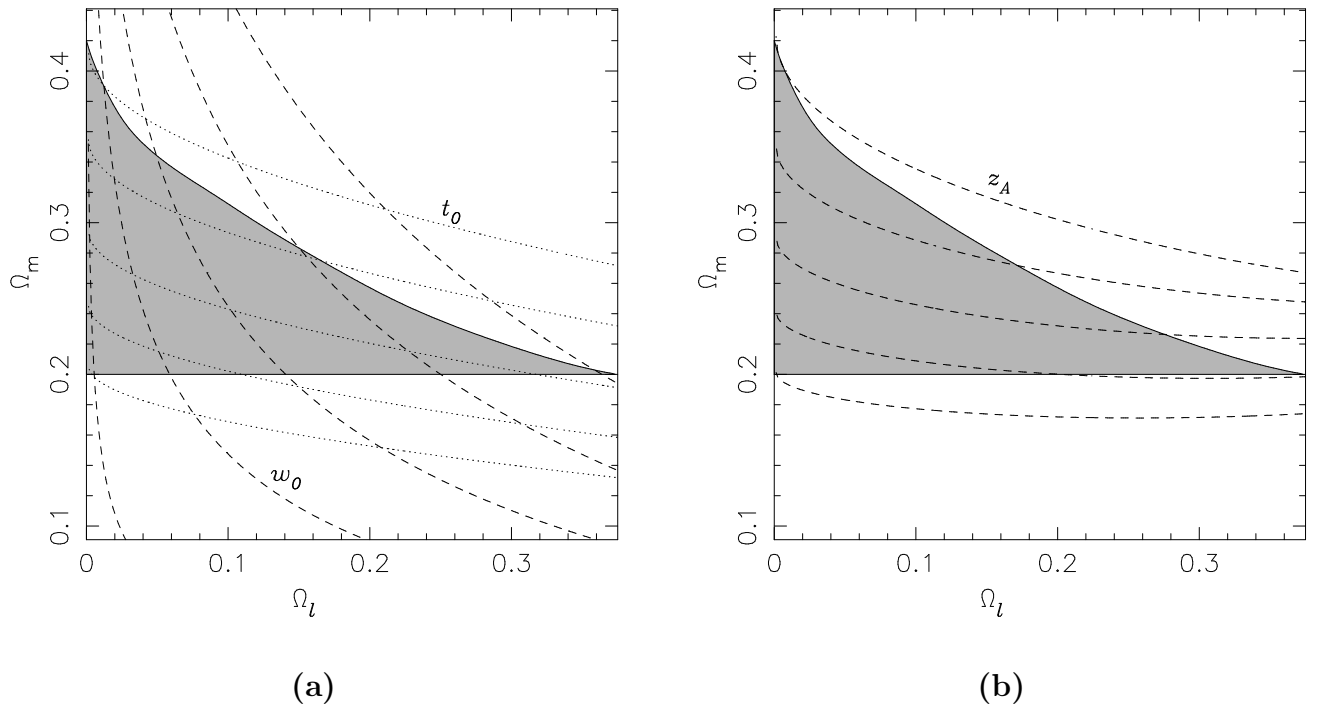

\begin{center}
$\begin{array}{c@{\hspace{0.4in}}c}
\multicolumn{1}{l}{\mbox{}} &
        \multicolumn{1}{l}{\mbox{}} \\ [-1.0cm]
\epsfxsize=3.2in
\epsffile{conf2_0.epsi} &
        \epsfxsize=3.2in
        \epsffile{conf2_z.epsi} \\ [0.20cm]
\mbox{\bf (a)} & \mbox{\bf (b)}
\end{array}$
\end{center}
\caption{\small
  The shaded region represents the intersection of the $3 \sigma$
  confidence level of BRANE2 for $\Omega_{\l_b}=0$ with the
  observational constraint $0.2 \leq \Omega_m \leq 0.5$. In (a) we
  show the values of different cosmological quantities determined at
  the present epoch. The dashed lines correspond to the current
  effective equation of state of braneworld dark energy for BRANE2
  models: $w_0 = -0.62, \ -0.75, \ -0.85,\ -0.92 \ {\rm and} \ -0.98$
  (top to bottom). The dotted lines correspond to the current age of
  the BRANE2 universe: $H_0 t_0 = 0.88, \ 0.92, \ 0.97, \ 1.02, \ {\rm
    and} \ 1.07$ from (top to bottom). This corresponds to $t_0({\rm
    Gyrs}) = 12.3, \ 12.9, \ 13.5, \ 14.3, \ {\rm and} \ 14.9$ if
  $H_0=70 \ {\rm km/sec/Mpc}$..  In (b) the dashed lines correspond to
  $z_A = 0.40, \ 0.55, \ 0.70, \ 0.85, \ {\rm and} \ 1.00$ (top to
  bottom), where $z_A$ is the epoch at which the braneworld universe
  (B2) begins to accelerate.  }
\label{fig:bounds2}
\end{figure*}

Fig. \ref{fig:contour2} shows the confidence levels in the
$\Omega_m-\Omega_l$ plane for the BRANE2 model.  Again the
intersection of the `$3\sigma$' SNe bound with the bound $0.2 \leq
\Omega_m \leq 0.5$ is shown as the dotted region. In contrast to
BRANE1 {\em lower values of $\Omega_m$ agree better with observations}
for BRANE2. The region to the right of the dashed line in the figure
corresponds to the class of braneworld models having future
singularities. It is interesting to note that such models are in good
agreement with SNe data for low values of $\Omega_m$, but are ruled
out (at $3\sigma$) if $\Omega_m \ggeq 0.25$.

Fig. \ref{fig:bounds2} shows lines of constant $w_0, ~t_0, {\rm and}
~z_A$ within the allowed region for BRANE2. Our results show that, the
age of the universe, its effective equation of state, and the
commencement of acceleration epoch, are constrained to lie within the
following intervals: $0.88 \lleq H_0t_0 \lleq 1.07$, $-1 \lleq w_0
\lleq -0.6$, and $0.4 \lleq z_A \lleq 1$. (One should note that $w_0
\leq -1$ in the BRANE2 model).

\begin{figure}
\centerline{ \psfig{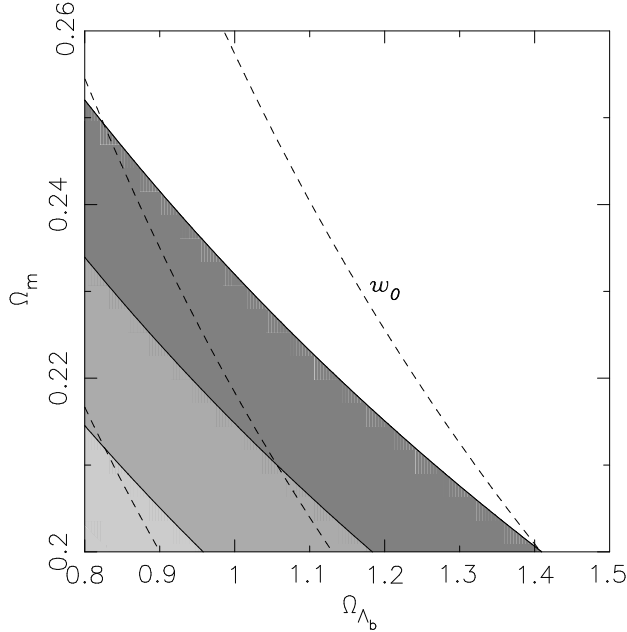} }
\bigskip
\caption{\small
  Confidence levels at $68.3\%$ (light grey inner contour) $95.4\%$
  (medium grey contour) and $99.73\%$ (dark grey outer contour) in the
  $\Omega_{\l_b}$-$\Omega_m$ plane for the `Disappearing Dark
  Energy' (DDE) model . The dashed lines show the current effective
  equation of state: $w_0 = -0.48, -0.44, -0.4$ (left to right).  
} \label{fig:contour3}
\end{figure}

Next we come to the third braneworld model, namely `Disappearing Dark
Energy' (DDE) for which the confidence levels in the
$\Omega_m-\Omega_{\l_b}$ plane are shown in figure \ref{fig:contour3}.
This case is different from the others in that the best fit value of
$\Omega_m$ tends to zero for this model.  This behaviour is easily
understood in view of the fact that the effective equation of state in
this model (\ref{eq:dde_w}) is always constrained to be $\ggeq -0.5$
and therefore cannot be very negative.  Since the luminosity distance
increases as $\Omega_m$ decreases, a sufficiently large value of the
luminosity distance (which agrees well with SNe observations) can only
be achieved at the expense of having a small value of $\Omega_m$ in
our braneworld model. Our results, shown in figure \ref{fig:contour3},
are for the DDE model constrained by the prior $\Omega_m \geq 0.2$.
The best fit DDE model ($\Omega_m = 0.2, ~ \Omega_{\l_b} = 0.8$) has a
chi-squared per degree of freedom of $\chi^2_{\rm dof} = 1.20$ which,
although larger than the value for LCDM ($\chi^2_{\rm dof} = 1.06 \ 
{\rm for} \ \Omega_m=0.3$, $\chi^2_{\rm dof} = 1.08 \ {\rm for} \ 
\Omega_m=0.2$), is much smaller than that for SCDM ($\chi^2_{\rm dof}
= 1.75$).

\begin{figure}
\centerline{ \psfig{figure=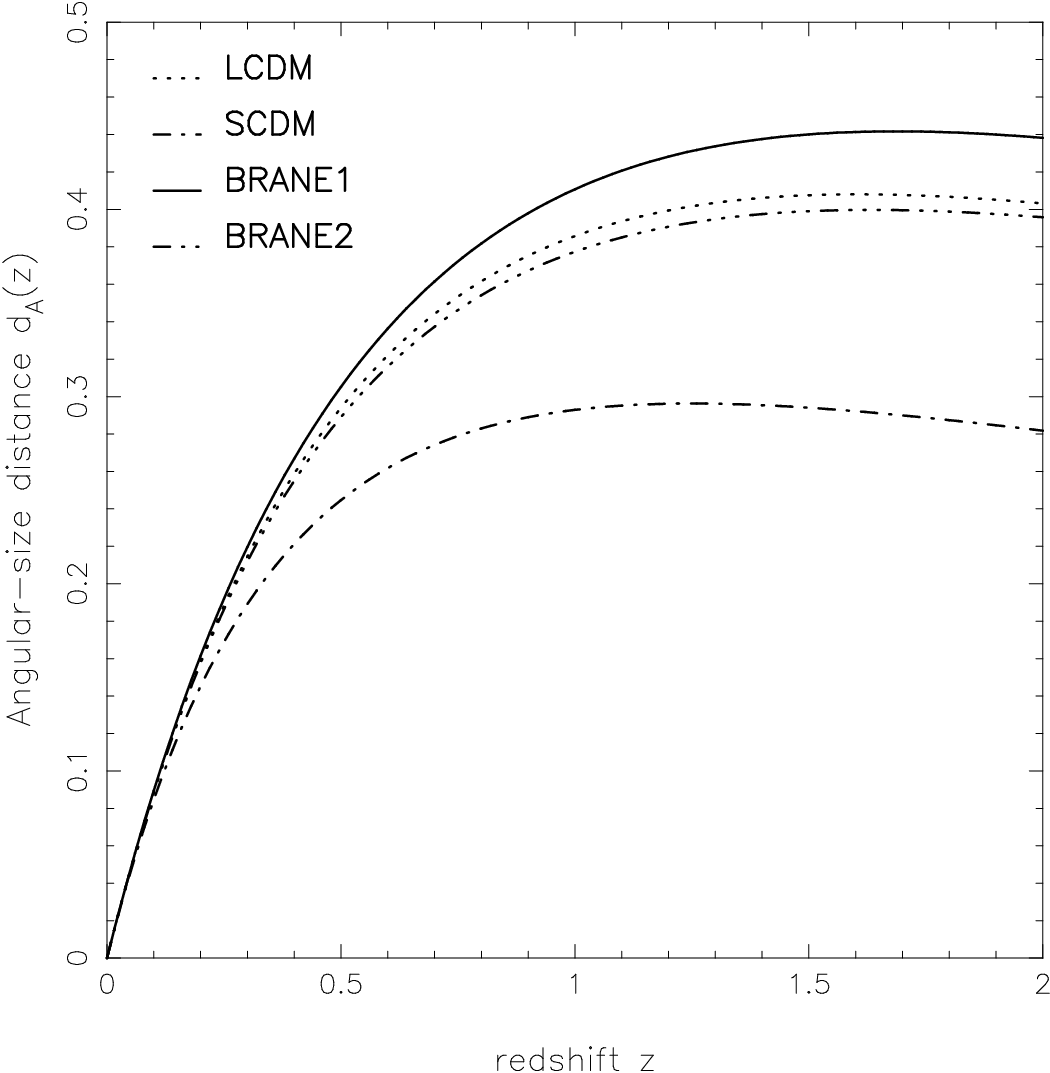,width=0.5\textwidth,angle=0} }
\bigskip
\caption{\small The angular-size distance (in units of $H_0^{-1}$)
in the braneworld models and
in LCDM, SCDM is shown as a function of the redshift.
The model parameters are:
BRANE1 ($\Omega_m = 0.3, \Omega_l = 0.3, \Omega_{\l_b} = 0$),
BRANE2 ($\Omega_m = 0.25, \Omega_l = 0.1, \Omega_{\l_b} = 0$),
LCDM ($\Omega_m = 0.3, \Omega_\l = 0.7$).
All three models satisfy existing supernovae bounds.
Also shown is the standard cold dark matter model (SCDM) with $\Omega_m = 1$.
}
\label{fig:angle}
\end{figure}

 \begin{figure}
 \centerline{ \psfig{figure=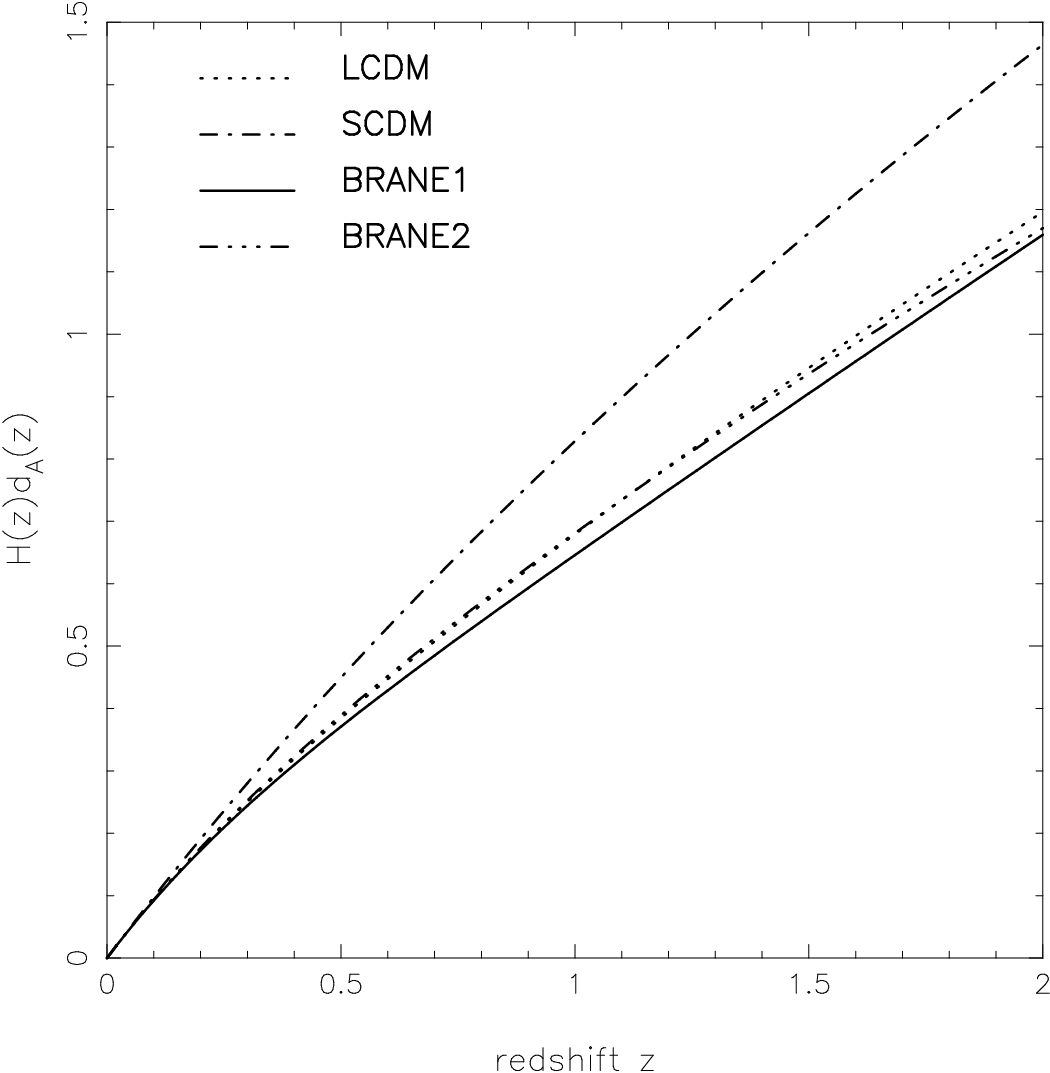,width=0.5\textwidth,angle=0} }
 \bigskip
 \caption{\small The dimensionless quantity $H(z)d_A(z)$ which plays
 an important role in the Alcock-Paczynski test is shown for
 braneworld models and
 for LCDM, SCDM as a function of the redshift.
 The model parameters are the same as in the previous figure.
 }
 \label{fig:alcock}
 \end{figure}

\begin{figure}
\centerline{ \psfig{figure=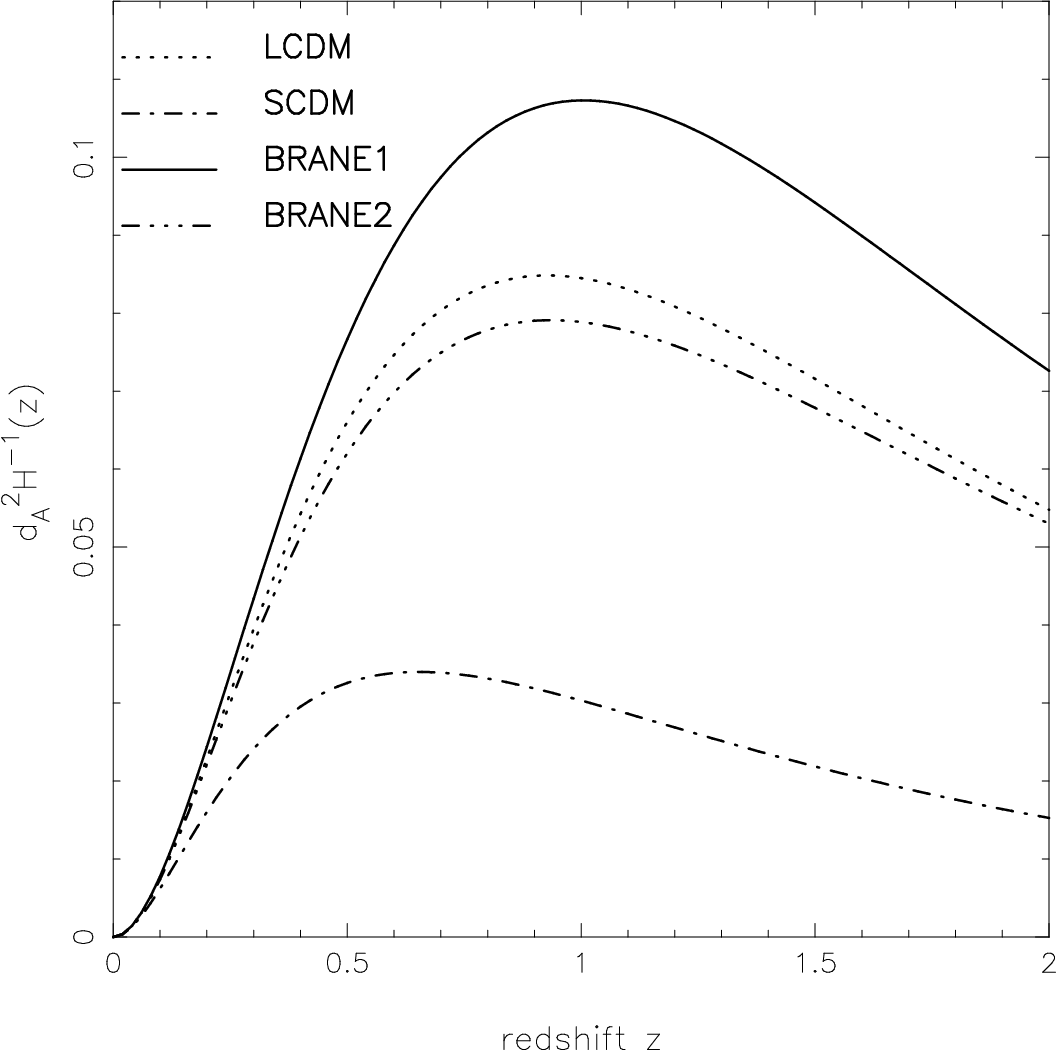,width=0.5\textwidth,angle=0} }
\bigskip
\caption{\small The quantity $d_A^2 H^{-1}(z)$ which plays a key
role in the volume-redshift test is shown for
braneworld models and
for LCDM, SCDM as a function of the redshift.
The model parameters are the same as in the previous figures.
}
\label{fig:volume}
\end{figure}

\begin{figure}
\centerline{ \psfig{figure=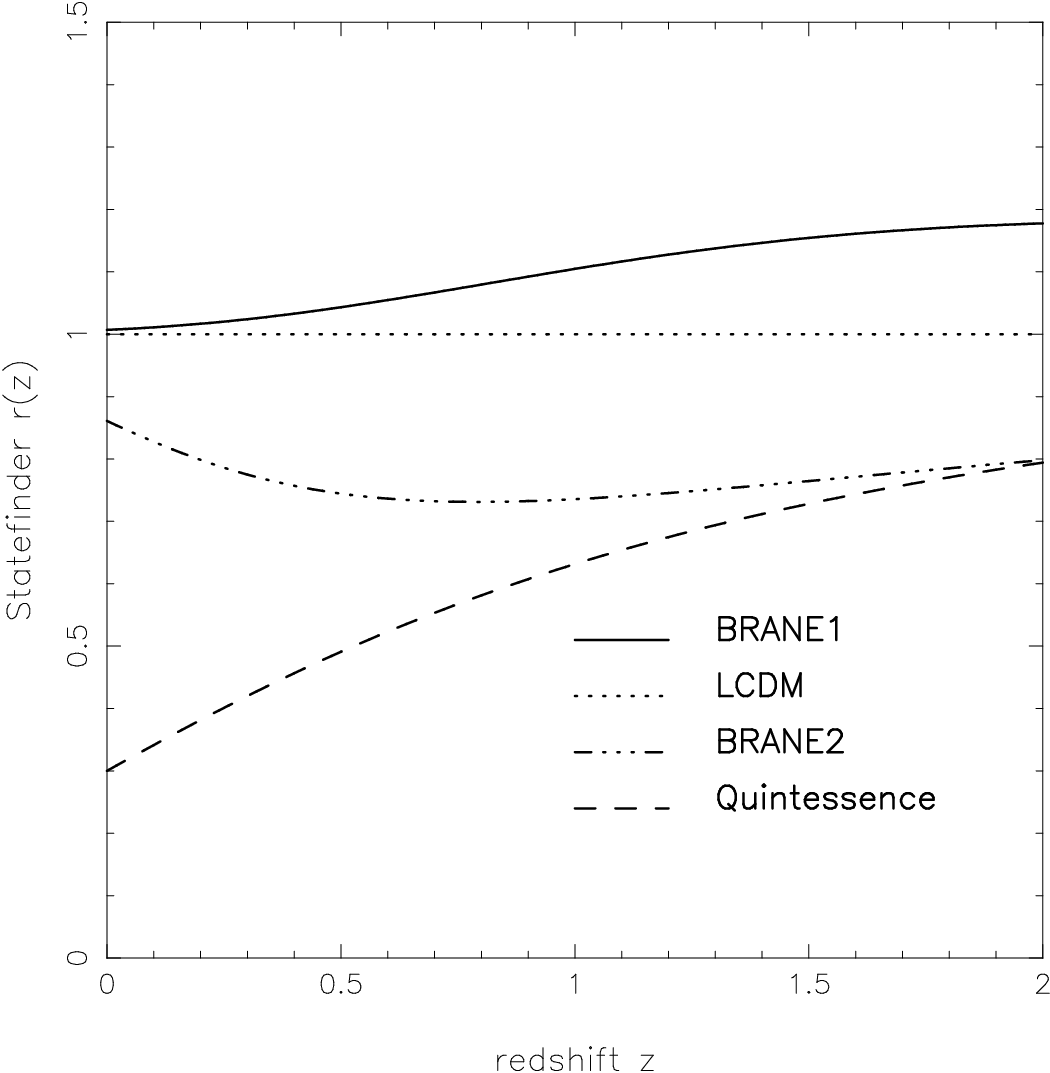,width=0.5\textwidth,angle=0} }
\bigskip
\caption{\small The dimensionless statefinder
$r = \,\,\stackrel{...}{a}/a H^3$ is shown for
braneworld models and
for LCDM. Also shown is the quintessence model with equation of state
$w = -2/3$.
The model parameters are the same as in the previous figure.
}
\label{fig:state1}
\end{figure}

\begin{figure}
\centerline{ \psfig{figure=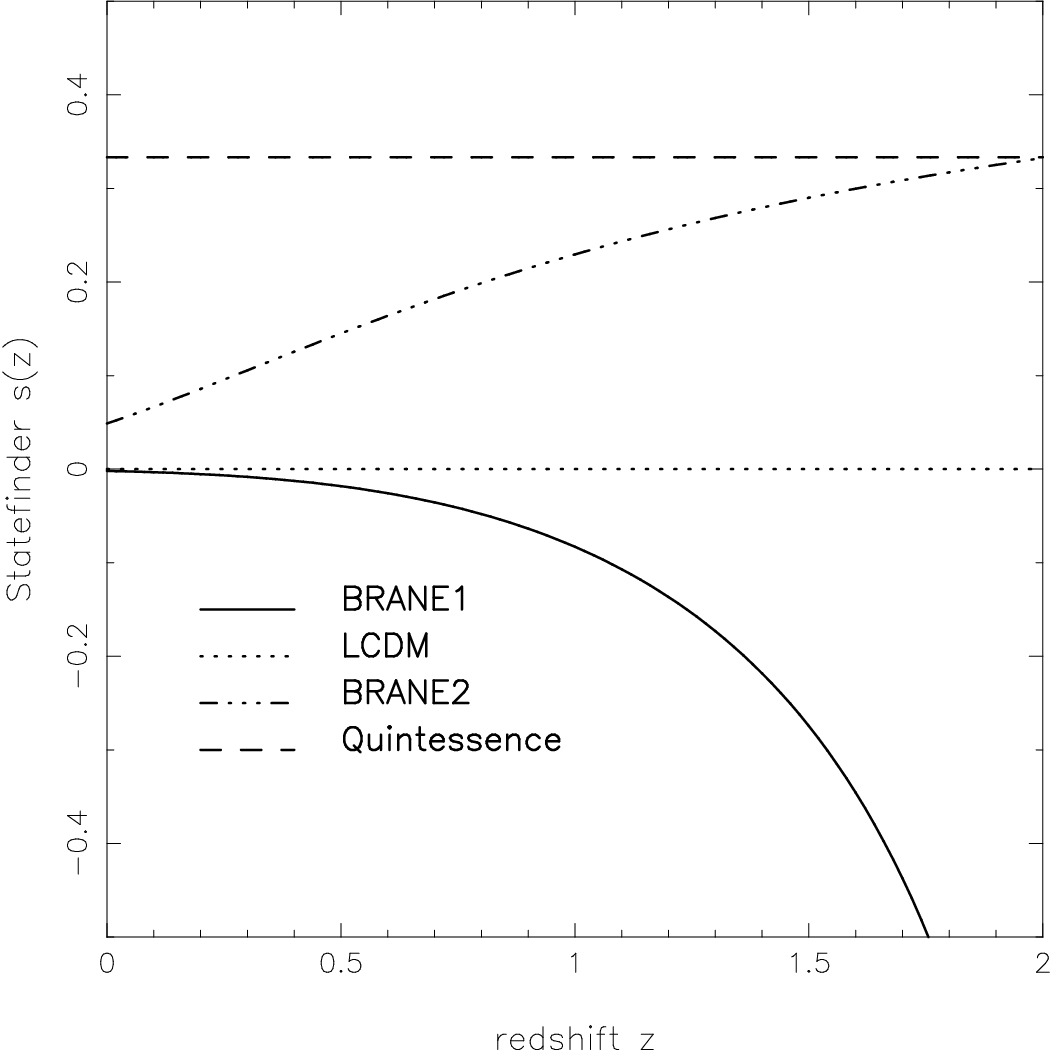,width=0.5\textwidth,angle=0} }
\bigskip
\caption{\small The dimensionless statefinder
  $s = (r - 1)/3(q - 1/2)$ is shown for braneworld models and for
  LCDM. Also shown is the quintessence model with equation of state $w
  = -2/3$.  The model parameters are the same as in the previous
  figure.  }
\label{fig:state2}
\end{figure}

One may try to constrain braneworld models further using some of the
cosmological tests outlined earlier in this paper. Through these tests
it may be possible to distinguish between different braneworld models
which agree well with current SNe data. To illustrate this we choose
two complementary braneworld models: (a) BRANE1 with $\Omega_m = 0.3,
\Omega_l = 0.3, \Omega_{\l_b} = 0$, and (b) BRANE2 with $\Omega_m =
0.25, \Omega_l = 0.1, \Omega_{\l_b} = 0$, which satisfy both the
high-$z$ SNe observations (at the $1\sigma$ level) and the
observational constraints on $\Omega_m$. In figure (\ref{fig:angle})
we show theoretical results for the angular-size distance $d_A$, which
is determined from observations of the angle $\Delta\theta$ subtended
by a `standard ruler' of length $\ell$ at a redshift $z$,
$\Delta\theta = \ell/d_A(z)$.  Another useful quantity is
$H(z)d_A(z)$, shown in figure (\ref{fig:alcock}), which plays a
crucial role in the Alcock-Paczynski anisotropy test. We clearly see
that the angular-size distance provides a better test with which
braneworld models may be distinguished from each other and from rival
models such as LCDM and SCDM. The Alcock-Paczynski test may not serve
this purpose, since both B1 \& B2 give results which are close to each
other and to those for LCDM.  The reason for this seeming degeneracy
lies in the fact that, for B1 (B2), the Hubble parameter $H(z)$ {\em
  decreases} (increases) relative to LCDM while the distance $d_A$
{\em increases} (decreases). Thus $H(z)$ and $d_A$ show opposing
tendencies for the braneworld models, which virtually neutralise each
other in the combination $H(z)d_A(z)$. Exactly the opposite effect is
achieved by the combination $d_A^2(z) H^{-1}(z)$ which plays a key
role in the volume-redshift test \cite{davis}. In this case the
difference between B1 and B2 becomes more pronounced as demonstrated
in figure \ref{fig:volume}.  We therefore conclude that the
volume-redshift test is probably a better means for differentiating
between the two braneworld models B1 \& B2 than the Alcock-Paczynski
test.

Interestingly the statefinder statistic (\ref{eq:state}) may also
provide us with a useful means by which to discriminate between rival
braneworld models and LCDM/quintessence, as demonstrated in figures
(\ref{fig:state1}) \& (\ref{fig:state2}). Other cosmological tests
which are likely to be useful in testing braneworld models include
gravitational lensing \cite{jain} and the cosmic microwave background
\cite{cmb}.  We shall return to these important issues in a future
work.

\section{Discussion and Conclusions}

This paper examines braneworld models of dark energy in the light of
recent supernova observations which indicate that the universe is
accelerating.  The braneworld models which we examine in this paper
have several interesting properties which distinguish them both from
the cosmological constant as well as from scalar field based `tracker'
models of dark energy. Like the latter, braneworld models presently
accelerate, and possess a longer age than the standard cold dark
matter model (SCDM).  However in marked contrast to both LCDM and
tracker models, the luminosity distance in one class of braneworld
models, B1, can be {\em greater} than the luminosity distance in LCDM
(for identical values of $\Omega_m$): $d_L^{~\rm dS}(z) \geq d_L^{~\rm
  BRANE1}(z) \geq d_L^{~\rm LCDM}(z)$, where $d_L^{~\rm dS}(z)$ refers
to the luminosity distance in spatially flat de Sitter space.  In
terms of the effective equation of state $w$, this is equivalent to
the assertion that $w \leq -1$.  This result is particularly
surprising since matter in the braneworld model never violates the
weak energy condition $\rho + p \geq 0$.  A maximum likelihood
analysis which compares braneworld model predictions with high
redshift type Ia supernovae data, shows that B1 models provide good
agreement with observations if $\Omega_m \ggeq 0.3$.  These results
broadly support the analysis of \cite{phantom} in which `phantom' dark
energy models, having the property $w=P/\rho \leq -1$, were compared
against supernova observations.  (It should be pointed out, however,
that `phantom models' invariably run into a physical singularity in
the future when $\rho_{\rm phantom} \to \infty$, such singularities
are absent in the B1 model which remains well behaved at all future
times.)

The second braneworld model we consider (B2) has properties which
complement those of B1, since $d_L^{~\rm LCDM}(z) \geq d_L^{~\rm
  BRANE2}(z) \geq d_L^{~\rm SCDM}(z)$.  This is equivalent to the
assertion that $-1 \leq w \leq 0$.  Results of a maximum likelihood
analysis show that B2 models are in excellent agreement with SNe data
for smaller values of the density parameter $\Omega_m \lleq 0.25$.

Finally braneworld models also permit the dark energy to be a
transient phenomenon. In models of this kind (called disappearing dark
energy: DDE) the acceleration of the universe takes place during a
transient regime separating past and future matter dominated epochs.
In these braneworld models, the universe does not possess an event
horizon and so it may be possible to reconcile a universe which
currently accelerates with the demands of string/M-theory.  Comparison
with Sne bounds shows that the Disappearing Dark Energy models
marginally satisfy existing supernova data provided $\Omega_m$ is
sufficiently small: $\Omega_m \lleq 0.23$. For larger values of
$\Omega_m$, this class of models may be on the verge of being ruled
out.

\section*{Acknowledgments}

We acknowledge useful discussions with Somak Raychaudhury, Tarun Deep
Saini, Yuri Shtanov, Tarun Souradeep and R.G. Vishwakarma. UA was
supported for this work by the CSIR.


\begin{thebibliography}{99}

\bibitem{kk} Appelquist, T., Chodos, A. and Freund, P.G.O. editors:
  {\sl Modern Kaluza-Klein Theories}, Addison-Wesley Publishing Co.
  (1987)

\bibitem{hw}
Horava, P. and Witten, E. Nucl. Phys.  {\bf B 460}, 606 (1996);
Nucl. Phys.  {\bf B 475}, 94 (1996).

\bibitem{rs}
L.~Randall and R.~Sundrum, Phys.\@ Rev.\@ Lett.\@ {\bf 83}, 3370 (1999) [{\tt
hep-ph/9905221}]; Phys.\@ Rev.\@ Lett.\@ {\bf 83}, 4690 (1999) [{\tt
hep-th/9906064}].

\bibitem{brane}
P.~Bin\'etruy, C.~Deffayet, and D.~Langlois, Nucl.\@ Phys.\@ B {\bf 565}, 269
(2000) [{\tt hep-th/9905012}]; \ C.~Cs\'aki, M.~Graesser, C.~Kolda, and
J.~Terning, Phys.\@ Lett.\@ B {\bf 462}, 34 (1999) [{\tt hep-ph/9906513}]; \
J.~M.~Cline, C.~Grojean, and G.~Servant, Phys.\@ Rev.\@ Lett.\@ {\bf 83}, 4245
(1999) [{\tt hep-ph/9906523}]; \ P.~Bin\'etruy, C.~Deffayet, U.~Ellwanger, and
D.~Langlois, Phys.\@ Lett.\@ B {\bf 477}, 285 (2000) [{\tt hep-th/9910219}].
\ T.~Shiromizu, K.~Maeda, and M.~Sasaki, Phys.\@ Rev.\@ D {\bf 62}, 024012 (2001)
 [{\tt hep-th/9910076}].
\ R.~Maartens, D.~Wands, B.~A.~Bassett, and I.~P.~C.~Heard, \prd {\bf 62}, 041301
  (2000) [{\tt hep-ph/9912464}]; \ E.~J.~Copeland, A.~R.~Liddle and J.~E.~Lidsey,
  \prd {\bf 64} 023509 (2001) [{\tt astro-ph/0006421}];
  \ V.~Sahni, M.~Sami, and T.~Souradeep, \prd {\bf 65} 023518 (2002) [{\tt gr-qc/0105121}].

\bibitem{anis}
\ R.~Maartens, V.~Sahni and T.~D.~Saini, \prd {\bf 63} 063509 (2001) [{\tt gr-qc/0011105}];
A.~Campos and C.~F.~Sopuerta, Phys.\@ Rev.\@ D {\bf 63}, 104012 (2001) [{\tt
hep-th/0101060}];
A.~V.~Toporensky, Class.\@ Quant.\@ Grav. {\bf 18}, 2311 (2001) [{\tt
gr-qc/0103093}].


\bibitem{dgp}
G.~Dvali, G.~Gabadadze, and M.~Porrati, Phys.\@ Lett.\@ B {\bf 485}, 208 (2000)
[{\tt hep-th/0005016}]; \ G.~Dvali and G.~Gabadadze, Phys.\@ Rev.\@ D {\bf 63},
065007 (2001) [{\tt hep-th/0008054}].

\bibitem{ss02a}
V.~Sahni and Yu.~V.~Shtanov, {\sl Braneworlds models of dark energy\/}, {\tt
astro-ph/0202346}.


\bibitem{riess}
A.~Riess \etal, Astron.\@ J.\@ {\bf 116}, 1009 (1998) [{\tt astro-ph/9805201}];

\bibitem{perl}
\ S.~J.~Perlmutter \etal, Astrophys.\@ J.\@ {\bf 517}, 565 (1999) [{\tt
astro-ph/9812133}].

\bibitem{ss00}
V.~Sahni and A.~A.~Starobinsky, Int.\@ J.\@ Mod.\@ Phys.\@ D {\bf 9}, 373
(2000) [{\tt astro-ph/9904398}].

\bibitem{sahni02}
V.~Sahni, Class. Quant. Grav. {\bf 19}, 3435 (2002)
[{\tt astro-ph/0202076}].

\bibitem{ss02b}
Yu.~V.~Shtanov and V.~Sahni, Class. Quant. Grav. {\bf 19}, L101 (2002)
[{\tt gr-qc/0204040}].

\bibitem{horizon}
W.~Fischler, A.~Kashani-Poor, R.~McNees, and S.~Paban, JHEP {\bf 0107}, 3
(2001) [{\tt hep-th/0104181}]; \ J.~Ellis, N.~E.~Mavromatos, and
D.~V.~Nanopoulos, {\sl String theory and an accelerating universe\/}, {\tt
hep-th/0105206}; \ J.~M.~Cline, JHEP {\bf 0108}, 35 (2001) [{\tt
hep-ph/0105251}]; \ X.-G.~He, {\sl Accelerating universe and event horizon\/},
{\tt astro-ph/0105005}.

\bibitem{CH}
H.~Collins and B.~Holdom, Phys.\@ Rev.\@ D {\bf 62}, 105009 (2000) [{\tt
hep-ph/0003173}].

\bibitem{Shtanov1}
Yu.~V.~Shtanov, {\sl On brane-world cosmology\/}, {\tt hep-th/0005193}.

\bibitem{DDG}
C.~Deffayet, G.~Dvali, and G.~Gabadadze, Phys.\@ Rev.\@ D {\bf 65},
044023 (2002) [{\tt astro-ph/0105068}];
C.~Deffayet,
S.~J.~Landau, J.~Raux, M.~Zaldarriaga, and P.~Astier, 
Phys.\@ Rev.\@ D {\bf 66}, 024019 (2002).
[{\tt astro-ph/0201164}].

\bibitem{sssa}
V.~Sahni, T. ~Saini, A.~A.~Starobinsky and U. ~Alam, {\sl Statefinder -- a new geometrical diagnostic of dark energy\/}, {\tt astro-ph/0201498}.

\bibitem{alcock}
C. Alcock and B. Paczynski, Nature {\bf 281}, 358 (1979).
J. Kujat, A.M. Linn, R.J. Scherrer and D.H. Weinberg,
ApJ {\bf S72}, 1 (2001) {\tt astro-ph/0112221}.

\bibitem{davis}
J.A. Newman and M. Davis, ApJ {\bf 534}, L11 (2000); astro-ph/0109131.

\bibitem{omegm}
J.~A. Peacock \etal, Nature {\bf 410}, 169 (2001).
J.~A.~Peacock \etal, {\sl Studying large-scale structure with the
2dF Galaxy Redshift Survey\/}, {\tt astro-ph/0204239 }.
To appear in "A New Era in Cosmology" (ASP Conference Proceedings), eds T. Shanks and N. Metcalfe.

\bibitem{borgani}
S. Borgani \etal, {\sl Measuring $\Omega_M$ with the ROSAT deep cluster
survey\/}, {\tt astro-ph/0106428}.

\bibitem{primack}
J. Primack, {\sl Status of Cold Dark Matter Cosmology\/},
{\tt astro-ph/0205391}. To appear in the
Proceedings of 5th International UCLA Symposium on Sources and Detection
of Dark Matter, Marina del Rey, February 2002, ed. D. Cline.

\bibitem{phantom}
R.~R.~Caldwell, {\sl A phantom menace?\/}, {\tt astro-ph/9908168}.

\bibitem{jain}
D. Jain, A. Dev and J.S. Alcaniz, {\sl Brane World Cosmologies and Statistical Properties of Gravitational Lenses\/}, {\tt astro-ph/0206224}.

\bibitem{cmb}
P.S. Corasanti and E.J. Copeland,
Phys.\@ Rev.\@ D {\bf 65}, 043004 (2002) [{\tt astro-ph/0107378}];
R. Bean and A. Melchiorri, Phys.\@ Rev.\@ D {\bf 65}, 041302 (2002)
[{\tt astro-ph/0110472}].

\end{thebibliography}
\end{document}